\documentstyle[aps,epsfig]{revtex}

\begin{document}

\title{Survival of orbiting in $^{20}$Ne (7 - 10 MeV/nucleon) + $^{12}$C reactions}

\author {C. Bhattacharya,  A. Dey,   S. Kundu, K. Banerjee, S. Bhattacharya,  S. Mukhopadhyay, D. Gupta\thanks {Present address : Institut de Physique Nucleaire, CNRS-IN2P3, 91406 Orsay cedex, France.}, T. Bhattacharjee, S. R. Banerjee, S. Bhattacharyya, T. Rana, S. K. Basu, R. Saha, S. Bhattacharjee,  K. Krishan\thanks {Present address : 306, VIP Enclave, VIP Road, Kolkata - 700 059, India}, A. Mukherjee\thanks {Present address : Saha Institute of Nuclear Physics, 1/AF Bidhan nagar, Kolkata - 700 064, India.}, D. Bandopadhyay\thanks {Present address : Department of Physics, University of Guelph, Guelph, ON Canada N1G 2W1.}}
\address{  Variable Energy Cyclotron Centre, Sector - 1, Block - AF,  Bidhan Nagar, Kolkata - 700 064, India.}
\author{C. Beck}
\address{ Institut de Recherches Subatomiques, UMR7500, CNRS-IN2P3 et Universite Louis Pasteur,
23, Rue du Loess, B.P. 28, F-67037, Strasbourg Cedex 2, France.}

\maketitle

\begin{abstract}
The  inclusive  energy  distributions  of  fragments  with Z~$\geq$~3
emitted from the bombardment of  $^{12}$C by $^{20}$Ne beams with
incident energies between 145 and 200 MeV  have
been  measured  in  the  angular  range  $\theta_{\rm lab}  \sim$ 10$^\circ$  -
50$^\circ$. Damped fragment yields in all cases have been found to
be characteristic of emission from fully energy equilibrated composites;
for B, C fragments, average Q-values, $<Q>$, were independent of the centre of mass  
emission angle ($\theta_{\rm c.m.}$),  
and  the  angular  distributions followed $\sim$1/sin$\theta_{\rm c.m.}$  
like variation, signifying long life times of  the  emitting  di-nuclear  systems. Estimation of
total yields of these fragments have been found to be much larger compared
to the standard statistical model predictions of the same. This may be indicative
of the survival of orbiting like process in $^{12}$C~+~$^{20}$Ne system at these
energies.

\end{abstract}


\noindent
Extensive efforts have been made in recent years to understand the mechanism of complex fragment emission 
in low-energy (E$_{\rm lab}$ $\lesssim$ 10 MeV/nucleon) light heavy-ion (A$_{\rm proj.}$~+~A$_{\rm targ.}$~$\lesssim$ 60) reactions \cite{Sanders99,bhat2002,bhat2004,bhat96,bec98,beck96,farrar96,shapi79,shapi821,shapi82,shiva} .  In most of the
reactions studied, the observed fully energy-damped  yields of the fragments
have been successfully explained in terms of  fusion-fission (FF) mechanism \cite{moretto,Sanders91,Matsuse97,dha2,Szanto97,Szanto96}. However, there has been a
noticeable exception  in the  reactions involving $\alpha$ - cluster nuclei (e.g. $^{20}$Ne + $^{12}$C, $^{28}$Si + $^{12}$C etc.), where large enhancement in yield and resonance-like excitation function in a few outgoing channels were observed, for which the deep-inelastic orbiting mechanism \cite{shapi79,shapi82,shiva} has been found to be particularly competitive. The first observation of fully energy damped reaction yields in light systems was reported by Shapira et. al. \cite{shapi79} in an investigation of $^{20}$Ne + $^{12}$C inelastic scattering  at backward angles, where large cross sections have been observed  in inelastic scattering yields near 180$^\circ$. Orbiting was then evidenced in $^{28}$Si + $^{12}$C reaction \cite{shapi82} and detailed studies on the same system revealed that, at lower bombarding energies, the excitation spectra for the   $^{12}$C  fragments were dominated by single excitation and mutual  excitations of the  $^{12}$C and $^{28}$Si fragments, whereas 
at higher bombarding energies,  the dominant strength for all these channels shifted to higher excitation energies \cite{shapi84}. For these higher energy spectra, the most probable Q-values were found to be independent of detection angle and the resulting angular distributions were found to have  d$\sigma$/d$\Omega \propto 1/sin{\theta_{\rm c.m.}}$ like angular dependence,  characteristic of a long-lived, orbiting,  dinuclear complex. Similar results have been obtained for $^{20}$Ne + $^{12}$C system \cite{shapi79,shapi821}, where resonance-like behaviour was also found in the excitation functions for several outgoing channels, similar to what has  been observed for symmetric   $^{16}$O + $^{16}$O system.  Subsequently, enhancements of large angle, binary reaction yields have been observed in somewhat heavier $^{28}$Si + $^{28}$Si, $^{24}$Mg + $^{24}$Mg systems \cite{Sanders99}, where significant non resonant background yield was observed at higher excitation energies.  All these are indicative of the fact that the enhancements are not due to specific structure effect appearing only in a few select channels; rather, they are manifestations of dynamics of damped nuclear reactions involving a large number of channels.

Intuitively, the enhancement in elastic and inelastic channels may be explained in terms of a long lived dinuclear configuration that decays back to entrance channel due to weak absorption which inhibits the orbiting configuration from spreading into compound nuclear states. However, weak absorption of only grazing partial waves is sufficient to explain the enhancement in the elastic channel; on the contrary, orbiting phenomenon in general suggests weak absorption for even lower partial waves between the critical angular momentum of fusion, $l_{\rm cr}$, and the  grazing angular momentum, $l_{\rm gr}$.  Furthermore, considering the rapid mass equilibration that is thought to occur in light systems, significant mass and charge transfer should also occur in course of evolution of the orbiting dinuclear complex. So, the rearrangement channels are also of interest in probing the dynamics of the orbiting process involving light nuclear systems. 

It is clear from the above that though some general consensus has been arrived at regarding the occurence of orbiting / resonance reactions (both are strongly correlated with number of open reaction channels, which in turn are related to weak absorption of partial waves ), precise mechanism(s) of orbiting and resonance behaviour is still unknown. For a better understanding of the orbiting process, it will be  interesting to study how the orbiting process evolves with energy. Intuitively, survival of long lived dinuclear configuration  other than fused composite is less probable at higher excitations and there are also indications  that entrance channel effect becomes smaller at higher energies \cite{Sanders99}. Shapira et. al. \cite{shapi821} made detailed study of $^{20}$Ne  + $^{12}$C system in the energy range E$_{\rm lab}$  =  54 - 81  MeV and showed that there was large enhancement of strongly damped yields characteristic of long lived orbiting $^{20}$Ne  + $^{12}$C dinuclear system. Our aim was to investigate whether the orbiting process would survive at even higher excitation energies, which might allow us to have a better understanding of orbiting vis-a-vis fusion-fission processes.  
 With this motivation,   we have studied the  fragment emission  spectra from the  reaction $^{20}$Ne  + $^{12}$C at E$_{\rm lab}$  =  145, 158, 168, 178 and  200  MeV, respectively in order to see whether orbiting still survives at higher energies.\\

\noindent
The  experiment was performed using  accelerated $^{20}$Ne  beams of energies 145, 158, 168, 178 and 200
 MeV, respectively, from the Variable Energy Cyclotron at Kolkata. The
target used was  550  $\mu$g/cm$^2$  self-supporting  $^{12}$C.  Different
fragments  have  been  detected using two solid state (Si(SB)) telescopes (
$\sim$ 10 $\mu$m $\Delta$E, 300 $\mu$m E) mounted in one arm  of  the  91.5  cm
scattering  chamber. Two solid state telescopes ($\sim$ 50, 100 $\mu$m $\Delta$E (Si(SB)) and 5 mm E(Si(Li)))  was mounted on the other arm  of  the  scattering  chamber for the detection of light charged particles; the same detectors were also used  as monitor detectors for normalisation purposes. Typical  solid  angle  subtended by each detector was
$\sim$0.3 msr. The telescopes were calibrated using elastically scattered $^{20}$Ne ion from Au, Al
and  Th-$\alpha$ source.  The  systematic  errors  in  the  data, arising  from  the uncertainties in the measurements of solid angle, target thickness and the calibration of current digitizer have been  estimated  to be  $\approx$  15\% (part of these uncertainties are due to the Gaussian fitting procedure employed to extract the fully damped yields from the sequential decay components).\\

\begin{figure} [h]

\vspace{.8cm}
\centering
{\epsfig{file=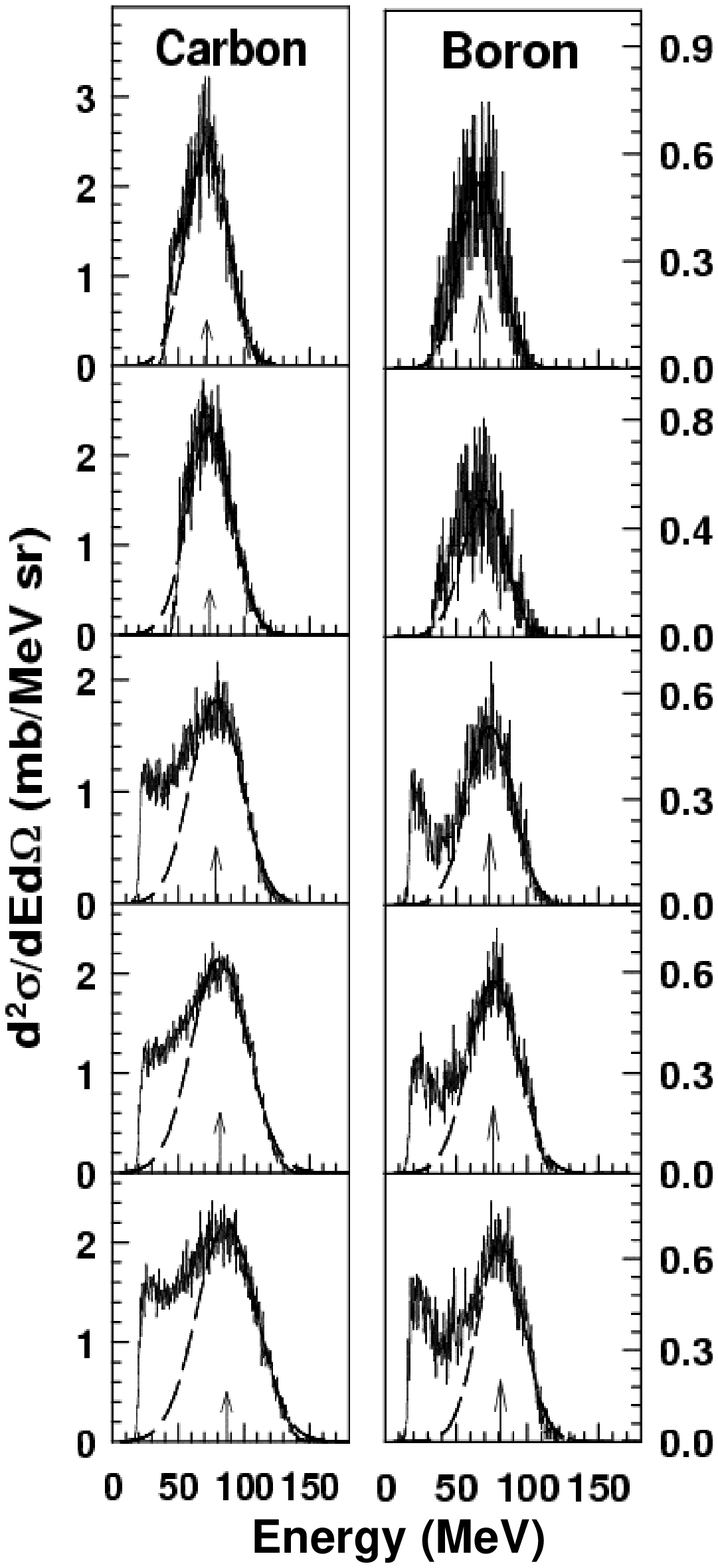,width=9.cm,height=9.8cm.}}

\vspace{.8cm}
\caption{ Measured energy distributions of the fragments B and C emitted  in the reaction $^{20}$Ne +$^{12}$C at E$_{\rm lab}$ = 145, 158, 168, 178  and 200 MeV, respectively (from top to bottom). The arrow corresponds to the position of the peak of the Gaussian distribution. } 
\label{nec1}
\end{figure}

\noindent
Inclusive  energy  distributions  for  various fragments (3$\leq Z \leq$12)
have been measured in the angular range 10$^\circ$-50$^\circ$. This covered backward
angles in the center of mass (c.m.) frame, because of the  inverse kinematics of the reactions. Typical energy spectra  for fragments  B and C obtained at an angle $10^{\circ}$ at E$_{\rm lab}$  =145, 158, 168, 178 and 200  MeV, respectively are shown in Fig.~\ref{nec1}. It is evident from Fig.~\ref{nec1} that energy spectra of the ejectiles  B and C at all bombarding energies exhibit strong peaking as a function of energy. In all cases, the energy distributions are nearly Gaussian in shape. 
The non-Gaussian shapes at the low-energy side of the energy spectra correspond to sequential decay processes which can be simulated by Monte Carlo statistical-model calculations as elaborated later in the discussion of data analysis.
The Gaussian fits so obtained are shown by dashed lines in Fig.~\ref{nec1}; the centroids (shown by arrow)  are found to correspond to  the scission of deformed dinuclear configuration  \cite{shiva,viola,beck96a}.  This suggests  that the fragments are emitted from fully energy relaxed composite - as expected for both FF and orbiting processes.  The increasing yields at lower energies may also be due to the second kinematical solution which is a  signature of binary nature of emission process.\\

\begin{figure} [h]

\vspace{.8cm}
\centering

{\epsfig{file=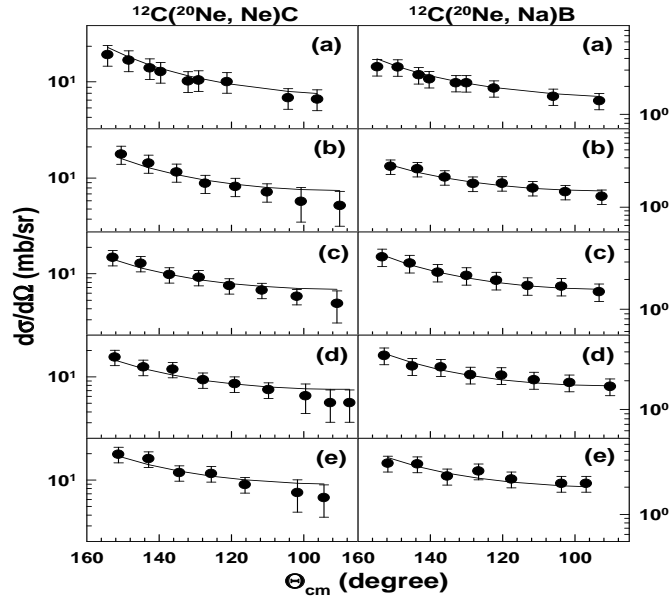,width=9.2cm,height=8.2cm}}

\vspace{.2cm}
\caption{The centre of mass angular distributions of the fragments B and C obtained at E$_{\rm lab}$  =145, 158, 168, 178  and 200 MeV, respectively (from a to e). Solid circles correspond to experimental data and the solid curves are f($\theta_{\rm c.m.}$)~$\sim$ ~1/sin$\theta_{\rm c.m.}$ fit to the data. }   
\label{nec2}
\end{figure}

\noindent
The centre of mass (c.m.) angular distributions of the fragments B and C obtained at 
 E$_{\rm lab}$  =145, 158, 168, 178  and 200 MeV, respectively, have been displayed as a function of centre of mass angle in Fig.~\ref{nec2}. The transformations from the laboratory to c.m. systems have been done with the assumption of a two body kinematics averaged over total kinetic energy distributions. 
It is seen that the c.m. angular distributions  of  these fragments obtained at all bombarding energies
follow the 1/sin$\theta_{\rm c.m.}$ like variation (shown by solid lines) - which further corroborate the conjecture of emission from fully equilibrated composite.\\

\begin{figure} [h]

\vspace{-.8cm}
\centering

{\epsfig{file=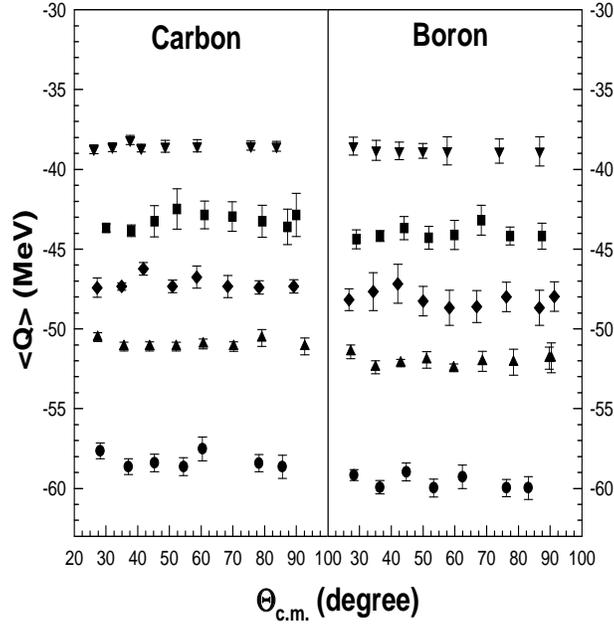,width=8.5cm,height=8.7cm}}

\vspace{.7cm}
\caption{Average   $<Q>$ values of the fragments B and C obtained at E$_{\rm lab}$  =145 , 158, 168, 178 and 200 MeV, (denoted by inverted triangle, square, diamond, triangle and circle, respectively)  plotted as a function of centre of mass emission angle.}   
\label{nec3}
\end{figure}

\noindent
The variations of average Q-value, $<Q>$, with centre of mass emission angle for the fragments B and C obtained at different E$_{\rm lab}$ have been shown in Fig.~\ref{nec3}. It is observed that for both the fragments at all energies, the average $<Q>$ values  are independent of the centre of mass emission angles. As the energy increases,  one expects that the contribution of deep inelastic process should start showing up. Such process is characterised by sharp fall of  $<Q>$ with angle, as observed in other light systems ($^{16}$O (116 MeV), $^{20}$Ne (145 MeV)  + $^{27}$Al,  $^{28}$Si systems, for example, where fragment energy distribution had two components, one coming from fully energy equilibrated fusion-fission, and the other from deep inelastic reactions \cite{bhat2002,bhat2004}). On the contrary,  for $^{20}$Ne  + $^{12}$C system, the  $<Q>$ values remain nearly constant which further suggest that at all angles, the fragments are emitted from completely equilibrated source  at all incident energies considered here. It is quite interesting to note that $^{16}$O, $^{20}$Ne  + $^{28}$Si systems, even though $\alpha$-like, do not fall in line with other light $\alpha$-like systems, like $^{20}$Ne  + $^{12}$C, $^{28}$Si  + $^{12}$C systems, for example, so far as the shape of the energy distributions, variation of  $<Q>$ values with angle,  and yields of the fragments are concerned. This difference in behaviour may be related to the number of open channels, which is much smaller in $^{20}$Ne  + $^{12}$C, $^{28}$Si  + $^{12}$C systems \cite{beck94}.

\begin{figure} [h]

\vspace{1.4cm}
\centering

{\epsfig{file=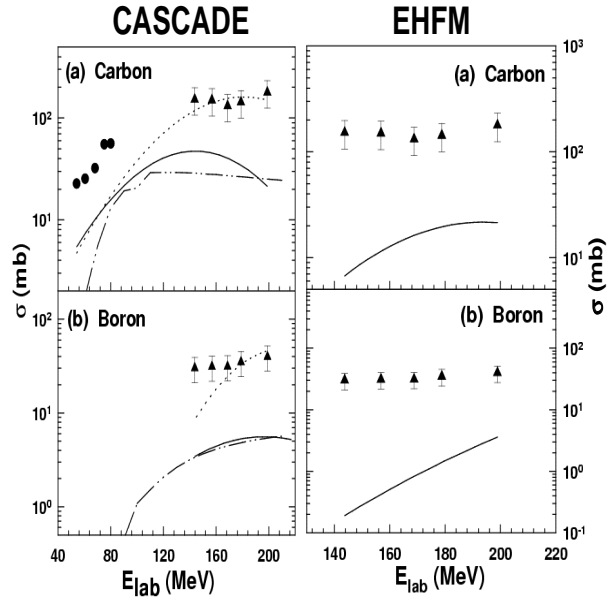, width=8.2cm,height=8.2cm}}

\vspace{.8cm}
\caption{The excitation functions for the angle integrated yield of the fragments Carbon and Boron. Triangles are the present data and lower energy data (shown by solid points) have been taken from \protect\cite{shapi79}. In the left side of the figure, solid, dashed and dash-dot-dotted curves are the predictions of the statistical model with $l = l_{\rm cr}$, $l_{\rm gr}$ and equilibrium orbiting model, respectively. Solid curves in the right side of the figure  are EHFM predictions.}   
\label{nec4}
\end{figure}

\noindent
It is clear from these observations that the yield of these fragments originates from fully energy relaxed events 
associated with the decay of either compound nucleus or long lived, orbiting dinuclear system. 
The possibility that the  fragments B, C  might be produced  wholly or partially through the  processes mentioned above   has been investigated more quantitatively by comparing the experimental yields with the theoretical predictions of the standard statistical model \cite{pul}, extended Hauser-Feshbach   model   (EHFM)   \cite{Matsuse97}. The solid circles and triangles in Fig.~\ref{nec4} are the experimental  angle integrated yields of the fragments. The solid lines 
in  Fig.~\ref{nec4} (left side) are the predictions of statistical model code CASCADE \cite{pul} . The calculations, considering  $l$ values up to $l_{\rm cr}$  at each energies ($l_{\rm cr} (\hbar)$ = 24, 24, 24, 25, 25 for E$_{\rm lab}$ = 145, 158, 168, 178, 200 MeV, respectively \cite{sb88}), are  found to underpredict the experimental yields.   Similar observation has been reported by Shapira {\it et al.} \cite{shapi79,shapi821} for the same system at lower energies.  
In Fig.~\ref{nec4} (right side), the EHFM predictions of the experimental yields of C and B  (solid lines) have been displayed alongwith the present experimental data (filled triangles). It is interesting to  find that the EHFM predictions are  also similar to those obtained from CASCADE calculations and 
the experimental yields are in fair excess of the   theoretical estimates of both CASCADE and EHFM. 

\begin{figure} [h]

\vspace{1.5cm}
\centering

{\epsfig{file=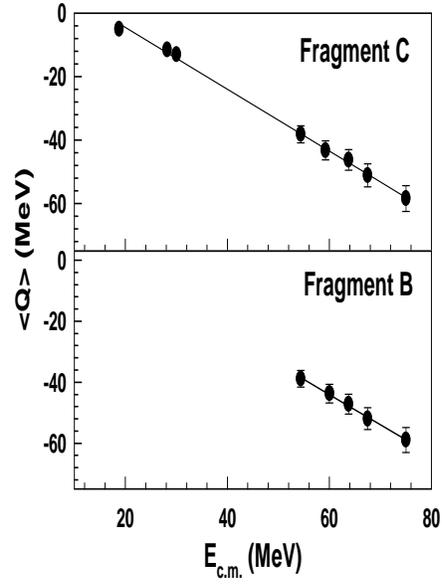,width=6.2cm,height=8.2cm}}

\vspace{.5cm}
\caption{Bombarding energy dependence of the average $<Q>$ values. The solid line shows the linear dependence of  $<Q>$  with bombarding energy. The  $<Q>$ values  at lower energies  are taken from \protect\cite{shapi821}.}   
\label{nec5}
\end{figure}

\noindent
Since the statistical models are not found to reproduce most of the observed experimental behaviors,
an additional reaction component corresponding to the orbiting mechanism has to be considered. The large
measured cross sections led to the suggestion that an orbiting, dinuclear configuration is
formed that decays back to the entrance channel. After the discovery of orbiting in the $^{12}$C+$^{28}$Si
system, similar enhancements of large-angle, binary-reaction yields are also observed in the present data.
It is expected that the orbiting mechanism will retain a greater memory of the entrance channel than the
fusion-fission process. The trapped, dinuclear complex can either evolve with complete amalgamation into a fully
equilibrated compound nucleus or, alternatively, escape into a binary exit channel by way
of orbiting trajectories.  Orbiting can therefore be described in terms of the formation
of a long-lived dinuclear molecular complex which acts as a ``doorway" state to fusion with a strong memory of the
entrance channel.  The equilibrium orbiting model has been used
to successfully explain both the observed cross sections and total kinetic energy (TKE) values of the fully damped fragments for
several lighter nuclear systems at lower energies. However in the present case, the theoretical predictions of the equilibrium orbiting model \cite{shiva} (shown in  Fig.~\ref{nec4} (left side) as dash-dot-dotted curves) are also found to 
underpredict the experimental yields. It is, therefore, evident that both the equilibrium orbiting and statistical decay (CASCADE, EHFM) models result in comparable disagreement with the data. It may be interesting to note here that Shapira {\it et al.} studied the same reaction at lower energies \cite{shapi79,shapi821} and came to the conclusion that the large enhancements in the energy damped fragment yield observed at those energies might be due to nuclear orbiting phenomenon. The average Q values ($<Q>$)  for the fragments C and B have been plotted in Fig.~\ref{nec5} as  function of  incident energy. The linear dependence of $<Q>$ with energy provides strong evidence that the long life time may be associated with an orbiting phenomenon. This linear dependence  of $<Q>$ can be expressed by the following simple equation, i.e., {$<Q>$= (14.9 $\pm$ 1.0) - (0.97 $\pm$ 0.02) E$_{\rm c.m.}$}.
It is interesting to note that the $<Q>$ values obtained in the present experiment between 145 MeV and 200 MeV fall on the same straight line extrapolated  from the lower energy ($\sim$50 - 80 MeV) data \cite{shapi821}. This means that the energy relaxation is complete for the fragment emission studied here up to the  incident energy of 200 MeV. Moreover, it also means that the final kinetic energy {(E$^f_{\rm kin}$ = $<Q>$ + E$_{\rm c.m.}$)} is nearly independent of bombarding energy - which may be due to the limitation on the maximum value of angular momentum beyond which the formation of  dinucleus is not allowed due to centrifugal repulsion \cite{shapi84}.\\

\noindent
In general, the energy distributions, the angular distributions and the total fragment yields measured for $^{20}$Ne + $^{12}$C reaction at incident energies between 145 MeV and 200 MeV are similar to those obtained at lower incident energies ($\sim$ 50 - 80 MeV) for the same system (see Refs.\cite{shapi79,shapi821}). Large energy damping, 1/sin$\theta_{\rm c.m.}$ dependence of angular distribution and near constancy of $<Q>$ over a wide angular range signify that the fragment decay originates from a long lived, fully energy equilibrated system. However, the large enhancement of fragment emission cross section over the statistical model predictions leads to the conjecture that the orbiting mechanism may still play a major role at these energies, though the equilibrium orbiting model of Ref. \cite{shiva} also fails to explain the large enhancement in yield. The dash-dot-dotted curve displayed in Fig.~\ref{nec4} (left side) represents the "best fit" that can be obtained by the orbiting model with a reasonable choice of the Bass potential parameters (strengths, short range, and long range of the proximity potential). To check whether the enhancement in C yield could be due to feeding from the secondary decay of heavier fragments of various possible binary break-up combinations, we have performed detailed simulations of secondary decay using the Monte carlo binary decay version of the statistical decay code LILITA \cite{lilita}. Secondary decay of Ne$^*$ (binary channel $^{20}$Ne + $^{12}$C), Mg$^*$ (binary channel $^{28}$Mg + $^{8}$Be), Si$^*$ (binary channel $^{28}$Si + $^{4}$He), and O$^*$ (binary channel $^{16}$O + $^{16}$O) have been studied. Indeed the LILITA calculations (using the parameter set proposed in the Appendix of Ref. \cite{shapi821} and assuming the excitation energy division follows the mass ratio \cite{Sanders99}) are in qualitative agreement with the experimental results obtained at 9 MeV/nucleon by Rae {\it et al.} \cite{Rae} for the sequential decay of $^{20}$Ne + $^{12}$C. It was found that even at the highest excitation energy, secondary decay of Mg$^*$ and Si$^*$ do not reach up to C; the contribution of primary Ne$^*$ decay to C yield was estimated to be $\sim$15--20\% of the primary yield. Nearly $\sim$40--45\% of the primary O$^*$ produced through binary exit channel$^{16}$O + $^{16}$O decays to C; however, as the O binary yield is small ($\sim$10\% of the binary Ne yield, as estimated from CASCADE \cite{pul}), overall secondary decay contribution from O is smaller than that from Ne. Moreover, the simulations of energy distributions of the secondary decay yield of C from Ne as well as O using the code LILITA show that they peak at much lower energies (typically, at $\sim 45 - 50$ MeV for Ne, and $\sim 55 - 60$ MeV for O, compared to the peak of the experimental energy distribution  at $\sim$ 70--90MeV). Now, the Gaussian fitting procedure for the extraction of primary C yield is fairly efficient in rejecting most of the low energy tail (typical rejection ratio $\sim$ 25--30 \% of the total yield). It is thus evident that the secondary decay component does not interfere with the estimated primary C yield for two reasons; firstly, total secondary decay yield is not quite large, and secondly, the Gaussian fitting procedure for the extraction of primary C yield does take care, to a large extent, of the rejection of the contributions of the secondary decay components as their energy distributions are different from those of the primary components.

\noindent
In conclusion, the present analysis seems to indicate that there is significant enhancement in fragment yield for $^{20}$Ne~+~$^{12}$C reactions in excess to the corresponding equilibrium model predictions \cite{shiva}. The shortcomings of the equilibrium model for orbiting does not imply that the presence of an orbiting mechanism, as distinct from fission, can be ruled out. On the contrary, there may be a large orbiting-like contribution from non fusion window ( $l_{\rm cr} \leq l \leq l_{\rm gr}$). This is consistent, at least qualitatively, with the fact that, CASCADE calculation \cite{pul} performed with $l$ values up to  $l_{\rm gr}$ ($l_{\rm gr} (\hbar)$ = 31, 33, 34, 36, 38 for E$_{\rm lab}$ = 145, 158, 168, 178, 200 MeV, respectively \cite{bass})  is found to reproduce the data fairly well ( dotted lines in Fig.~\ref{nec4} (left side)). Yields in the transfer channels (B, for example) are also found to be strongly affected by the orbiting process (yield enhancement), which may be due to stochastic nucleon exchanges during long lifetime of the dinuclear system. It is interesting to mention that, $^{16}$O, $^{20}$Ne  + $^{28}$Si systems, even though $\alpha$-like, do not show the characteristics of orbiting at these energies, but orbiting-like behaviour has been observed for $^{28}$Si  + $^{16}$O reaction at lower energies \cite{Szanto97}. Thus, the enhancement in fragment yield observed in the present data is primarily believed to be due to orbiting phenomenon; however,the equilibrium orbiting model, in its present form, seems to be inadequate to explain the whole set of data, and a more complete understanding of orbiting and vis-a-vis the angular momentum dissipation (which plays a crucial role indefining orbiting trajectories and yield) will be required.\\

\noindent

\acknowledgements
The authors like to thank the cyclotron operating crew for smooth running of the machine, and H. P. Sil for the fabrication of thin silicon detectors for the experiment. One of the authors ( A. D. )  acknowledges with thanks the financial support provided by the Council of Scientific and Industrial Research, Government of India.

\newpage

\newpage

\newpage

\end{document}